\documentclass[%
 aip,
nofootinbib,
 amsmath,amssymb,
 reprint,%
]{revtex4-1}
\usepackage{graphicx,epsf,epsfig}
\usepackage{subfigure}
\usepackage{dcolumn}
\usepackage{bm}
\usepackage{textcomp}
\usepackage{amsmath}
\usepackage[normalem]{ulem} 
\usepackage{tikz}
\usepackage{multirow}
\usepackage[utf8]{inputenc}
\usepackage[ngerman,english]{babel}
\usepackage{tikz}
\usepackage{url}
\usepackage{nomencl} 
 \usepackage{framed} 
\usepackage{etoolbox}
\usepackage{color}
\usepackage{amsmath}
\usepackage{subfigure}
\renewcommand\nomgroup[1]{%
  \item[\bfseries
  \ifstrequal{#1}{S}{Variables}{%
  \ifstrequal{#1}{A}{Abbreviations}{%
  \ifstrequal{#1}{F}{Fundamental constants}{%
  \ifstrequal{#1}{G}{Greek Symbols}{}}}}%
]}

\begin{document}

\title{Carrier transport and performance limit of semi-transparent photovoltaics: CuIn$_{1-x}$Ga$_x$Se$_2$ as a case study}

\author{Eymana Maria}
\affiliation{Department of Electrical Engineering and Computer Science, University of Michigan, Ann Arbor, Michigan 48109, USA.}
\affiliation{Department of Electrical and Electronic Engineering, Bangladesh University of Engineering and Technology, Dhaka-1000, Bangladesh.}
\author{Ajanta Saha}
\affiliation{School of Electrical and Computer Engineering, Purdue University, West Lafayette, IN 47906 USA.}
\affiliation{Department of Electrical and Electronic Engineering, Bangladesh University of Engineering and Technology, Dhaka-1000, Bangladesh.}
\author{M. Ryyan Khan}
\affiliation{Department of Electrical and Electronic Engineering, East West University, Dhaka-1212, Bangladesh.}
\author{Md. Abdullah Zubair}
\email{rnsajjad@gce.buet.ac.bd, mazubair2017@gmail.com}
\affiliation{Department of Glass and Ceramic Engineering, Bangladesh University of Engineering and Technology, Dhaka-1000, Bangladesh.}
\author{Md. Zunaid Baten}
\affiliation{Department of Electrical and Electronic Engineering, Bangladesh University of Engineering and Technology, Dhaka-1000, Bangladesh.}
\author{Redwan N. Sajjad}
\email{rnsajjad@gce.buet.ac.bd, mazubair2017@gmail.com}
\affiliation{Department of Glass and Ceramic Engineering, Bangladesh University of Engineering and Technology, Dhaka-1000, Bangladesh.}


\begin{abstract}\label{abstract}
Semi-transparent photovoltaic devices for building integrated applications have the potential to provide simultaneous power generation and natural light penetration. CuIn$_{1-x}$Ga$_x$Se$_2$ (CIGS) has been established as a mature technology for thin-film photovoltaics, however, its potential for Semi-Transparent Photovoltaics (STPV) is yet to be explored. In this paper, we present its carrier transport physics explaining the trend seen in recently published experiments. STPV requires deposition of films of only a few hundred nanometers to make them transparent and manifests several unique properties compared to a conventional thin-film solar cell. Our analysis shows that the short-circuit current, $J_\mathrm{sc}$ is dominated by carriers generated in the depletion region, making it nearly independent of bulk and back-surface recombination. The bulk recombination, which limits the open-circuit voltage $V_\mathrm{oc}$, appears to be higher than usual attributable to numerous grain boundaries. When the absorber layer is reduced below 500 nm, grain size reduces resulting in more grain boundaries and higher resistance. This produces an inverse relationship between series resistance and absorber thickness. We also present a thickness-dependent model of shunt resistance showing its impact in these ultra-thin devices. For various scenarios of bulk and interface recombinations, shunt and series resistances, $AVT$ and composition of CuIn$_{1-x}$Ga$_x$Se$_2$, we project the efficiency limit which - for most practical cases -  is found to be $\leq$10\% for $AVT\geq$25\%.  
\end{abstract}
\maketitle



\section{Introduction}
Photovoltaic (PV) technology is a major hope for the world to combat climate change and satisfy the ever increasing energy demand by harnessing energy from the sun. Over the past several decades, research on PV technology has produced several generations of solar cells  - crystalline and polycrystalline silicon solar cells, thin-film solar cells and the third generation solar cells based on novel materials. One of the roadblocks to generate large-scale electricity from PV is often the lack of adequate space, which hinders its installation in urban areas and countries with dense populations. Semi-Transparent Photovoltaics (STPV) \cite{traverse2017emergence} - an emerging variant of the conventional thin-film PV technology - offers a way to generate power without creating additional footprint. When installed on building windows, STPV can allow visible light and at the same time produce electricity for buildings which is a major electricity consuming sector. Depending on location,  about 45-90\% building energy demand can be satisfied with STPV \cite{refat2020prospect}, making it a promising technology to realize net-zero energy building\footnote{The article has been submitted to the \textit{Journal of Applied Physics}. After it is accepted, it will be found at ``https://aip.scitation.org/journal/jap"}.    

While thin-film PV technology has been under research for many years, research on STPV is still in its early stage. One simple approach to fabricate STPV is to reduce the absorber layer thickness and replace the back contact with a transparent electrode, $e.g.$ Indium Tin Oxide (ITO). The required absorber thickness depends on its absorption coefficient and - for high absorption materials such as CuIn$_{1-x}$Ga$_x$Se$_2$ (CIGS) - it is found to be just a couple of hundreds of nanometer to achieve about 25\% transparency \cite{saifullah2016development}. The reduction of the absorber thickness to such a length scale creates many challenges $e.g.$ high bulk recombination and shunt conduction. However, what makes it interesting at the same time is that the carrier transport and the dependence of output characteristics on material parameters are quite different at this length scale, such as their dependence on bulk defects, interface and back surface recombination. The subject matter of this paper is to investigate these unique transport properties and identify engineering options for maximum efficiency. 
 
Several materials and device designs have been reported so far to realize STPV such as wavelength selective STPV based on organic materials \cite{chen2012visibly, wang2017fused}, non-wavelength selective STPV based on amorphous silicon thin-film \cite{chae2014building}, crystalline silicon \cite{kuhn1999characterization} and perovskites \cite{kwon2016parallelized, chang2015high}. CuIn$_{1-x}$Ga$_x$Se$_2$ (CIGS) based thin-film solar cells are reported to have one of the highest efficiency (23.35\%) among all thin-film technologies \cite{eff_table}. So it is natural to inquire how it will perform as STPV. In the recent past, fabrication of CIGS based non-wavelength selective STPV \cite{saifullah2016development, kim2020flexible, shin2019semi,cavallari2017low} and tandem structures with perovskite \cite{paetzold2017scalable, guchhait2017over, bailie2015semi} have been reported. There have been few theoretical studies on STPV focusing on detailed balance limit of efficiency for STPV in general \cite{wheeler2019detailed}, figure of merit of performance \cite{lunt2012theoretical,treml2016quantitative} and color appearance of STPV \cite{forberich2015efficiency,lynn2012color}. There has been no comprehensive study on carrier transport in STPV and what limits transparency and efficiency in practical devices. With CIGS as the absorber material, our work illustrates the unique challenges and prospects of this new class of photovoltaic devices, keeping in mind the trade-off between transparency and efficiency. 
Our contributions in this paper are the following, 

1) We report the thermodynamic limit of efficiency of CIGS-based STPV and identify the mechanisms that limit the performance of the devices reported in experiments (sections III, IV). The detailed balance calculation shows efficiency dependence on transparency and electronic bandgap. The optimum bandgap of STPV depends on transparency and, depending on transparency, it can be different from opaque solar cells. We find that the performance-limiting factors in reported experiments \cite{saifullah2016development, kim2020flexible} are strong bulk recombination limiting the open-circuit voltage ($V_\mathrm{oc}$), high shunt conduction and series resistance originating from numerous grains in the thin absorber.  

2) We explain the underlying carrier transport mechanisms of CIGS based STPV vs that in conventional (opaque) devices (section V). We show that the dependence of $V_\mathrm{oc}$ and short-circuit current ($J_\mathrm{sc}$) on bulk, interface and back-surface recombination are different producing unique results.  
 
3) We project efficiency limits for various compositions of CuIn$_{1-x}$Ga$_x$Se$_2$ based on improved material parameters (section VI). In this regard, we show the change of $J_\mathrm{sc}$, $V_\mathrm{oc}$ with material parameters and present thickness dependent models for shunt ($R_\mathrm{sh}$) and series ($R_s$) resistance. We show that the reduction of film thickness leads to a rapid increase in grain boundaries leading to high series resistance. 
Resistances from these models (Section VI) are in good agreement with the values found while fitting the experimental $J$-$V$ (Section IV) - this consolidates our overall analysis. The increase of series resistance with reduced CIGS thickness reported in multiple experiments is also explained with the model. Performance projection is done for two different transparencies - 25\% and 40\%. Above 40\% $AVT$, efficiency falls below $\sim$5\% which may not be enough to reduce overall building energy consumption and below 25\% $AVT$, natural light penetration will be insignificant for most places \cite{refat2020prospect}.


\section{Optical and carrier transport model}

We use finite Difference Time Domain (FDTD) method to calculate optical reflection, absorption and transmission through the thin-film device. A coupled 2D Poisson and drift-diffusion solver is employed to calculate  device potential profile and carrier transport properties. Bulk recombination lifetime ($\tau$), CIGS carrier density ($N_A$), series ($R_s$), and shunt ($R_\mathrm{sh}$) resistances are used as fitting parameters while matching the experimental data.  Simulation results are discussed in relation to analytical formalism throughout the paper to provide physical insight. 

\section{Detailed balance limit of efficiency}
\begin{figure}[ht!]
\centering
\includegraphics[width=2.7in]{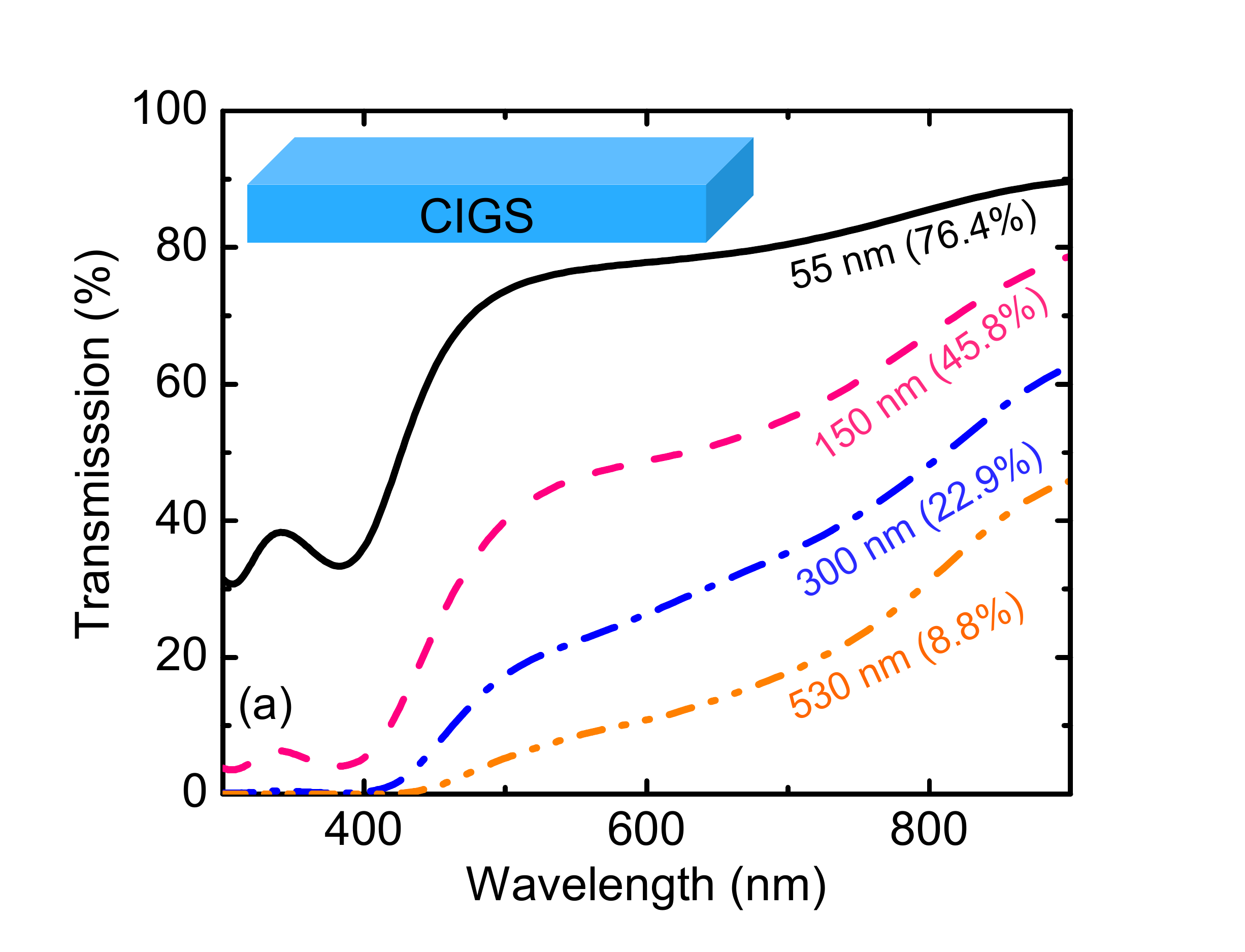}\quad
{\includegraphics[width=2.7in]{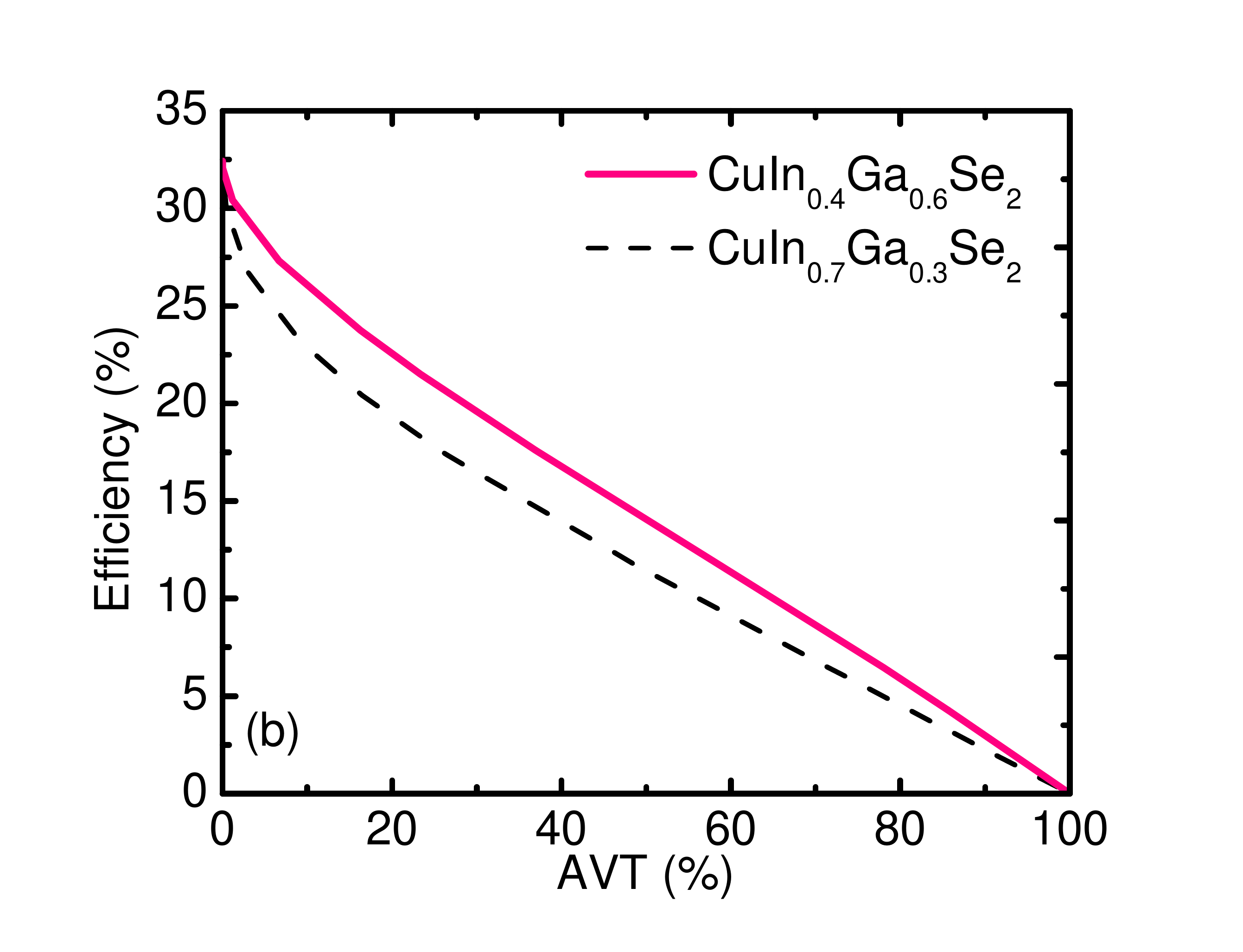}\quad}
\caption{(a) Transmission through a bare CIGS layer ignoring reflection for four different thicknesses. $AVT$ for each layer is shown in parenthesis. (b) Efficiency based on detail balance limit as a function of $AVT$ for two different gallium mole fractions in CuIn$_{1-x}$Ga$_x$Se$_2$. Efficiency is compromised as the material becomes more transparent (with reduced thickness). Higher gallium fraction increases bandgap and efficiency. } \label{fig1}
\end{figure}
The Shockley-Queisser limit of solar cell efficiency \cite{shockley1961detailed} is obtained by assuming that the absorber behaves as an ideal black body absorbing all incident electromagnetic waves with energy above its bandgap $E_g$. For STPV, not all incident waves above $E_g$ are absorbed; the absorption depends on the thickness. According to Beer-Lambert law, the thickness-dependent partial absorption is given by, ${Ab}(\lambda) = 1-e^{-\alpha t}$ with transmission, ${Tr}(\lambda) = e^{-\alpha t}$ for absorption coefficient $\alpha$ \cite{paulson2003optical} and film thickness $t$ at wavelength $\lambda$. Similar results are obtained from FDTD simulation as shown in Fig. \ref{fig1}a for four different CIGS thicknesses. For a given thickness, the transmission is higher at longer wavelengths due to lower absorption.  
%

To find the thermodynamic limit of efficiency, we vary CIGS layer thickness from 20 nm to 2000 nm and calculate transmission through and absorption in the layer ignoring reflection at the air-CIGS interface. The absorption probability ${Ab}$ from simulation is used to calculate the short-circuit current density, $J_\mathrm{sc}$ given by
\begin{eqnarray}
J_\mathrm{sc} = q\int_0^{\lambda_g}{\mathcal{S}}(\lambda)  {Ab}(\lambda)d\lambda
\end{eqnarray}
where ${\mathcal{S}}(\lambda)$ is the AM1.5G photon flux (/m$^2$-nm), ${\lambda_g}=hc/E_g$ with Planck’s constant $h$ and speed of light in vacuum $c$. 
Open circuit voltage is calculated using $V_\mathrm{oc}=\big({k_BT}/{q}\big)\mathrm{ln}\big({J_\mathrm{sc}}/{J_0}+1\big)$, where $J_0$ is the reverse saturation current under dark condition calculated as $J_0 = 9.66\times 10^{-15}$ mA/cm$^2$ (for $E_g$ = 1.18 eV, $x$ = 0.3) and $J_0 = 1.31\times 10^{-18}$ mA/cm$^2$ (for $E_g$ = 1.38 eV, $x$ = 0.6) from the expression given in \cite{green1982solar}, $k_B$ is the Boltzmann constant and $T$ is room temperature. The fill factor ($FF$) is found from Green's semi-empirical model \cite{green1982solar} given by, $FF=\big[{v_\mathrm{oc}-\mathrm{ln}(v_\mathrm{oc}+0.72)}\big]/({v_\mathrm{oc}+1})$, where $v_\mathrm{oc}={V_\mathrm{oc}}/({{k_BT}/{q}})$. The maximum efficiency, $\eta$ is given by
\begin{eqnarray}
\eta=\frac{V_\mathrm{oc}J_\mathrm{sc}FF}{P_\mathrm{in}}
\end{eqnarray}
where $P\mathrm{_{in}} = \int \mathcal{S}(\lambda) (hc/\lambda) d\lambda \approx$ 1000 W/m$^2$ is the incident solar power density for AM1.5G spectrum. 

Average Visible Transmittance ($AVT$) is calculated from transmission, ${Tr}$\cite{traverse2017emergence}
\begin{eqnarray}
AVT =  \frac{\int{Tr}(\lambda)\mathcal{P}(\lambda)\mathcal{S}(\lambda)d\lambda}{\int\mathcal{P}(\lambda)\mathcal{S}(\lambda)d\lambda}
\end{eqnarray} 
where $\mathcal{P}$ is the photopic response of human eye, non-zero only in the range of 300-700 nm. The Shockley-Queisser limit of efficiency, which is $\sim$ 31\% for $E_g \sim 1.4$ eV,  is captured by the detailed balance model for 0\% $AVT$ (Fig. \ref{fig1}b) at large thickness limit. As device thickness is decreased, efficiency decreases sharply for low $AVT$ due to lack of transmission in the visible range. $AVT$ starts increasing above a thickness threshold when transmission occurs in the visible range. For $AVT>10\%$, efficiency drops almost linearly with $AVT$, reaching 0 at 100\% $AVT$ at $\sim$ 20 nm thickness. It is noteworthy that bandgap dependence of efficiency for STPV is different from an opaque solar cell and depends on $AVT$ \cite{wheeler2019detailed}. We find the bandgap dependence of efficiency for CIGS for two gallium mole fractions - $x$ = 0.31 and $x$ = 0.66 having bandgaps of 1.18 eV and 1.42 eV respectively given by the following \cite{bar2004determination}
\begin{eqnarray}
E_g = 0.13x^2+0.55x+1
\end{eqnarray}
Thus CuIn$_{1-x}$Ga$_x$Se$_2$ with the higher gallium mole fraction ($x$) yields higher efficiency. However, the thermodynamic limit does not include details of carrier dynamics, series and shunt resistance. In the rest of the paper, we present experimental fitting and practical limit to efficiency (sections IV, V, VI) for $x$ =  0.71 with Bandgap of $E_g = 1.5$ eV reported in Ref. \cite{saifullah2016development} followed by the calculation of efficiency as a function of gallium mole fraction in section VI. 

\section{Comparison of the transport model with experiment}
%

%
%
\begin{figure}[ht!]
\centering
\includegraphics[width=2.2in]{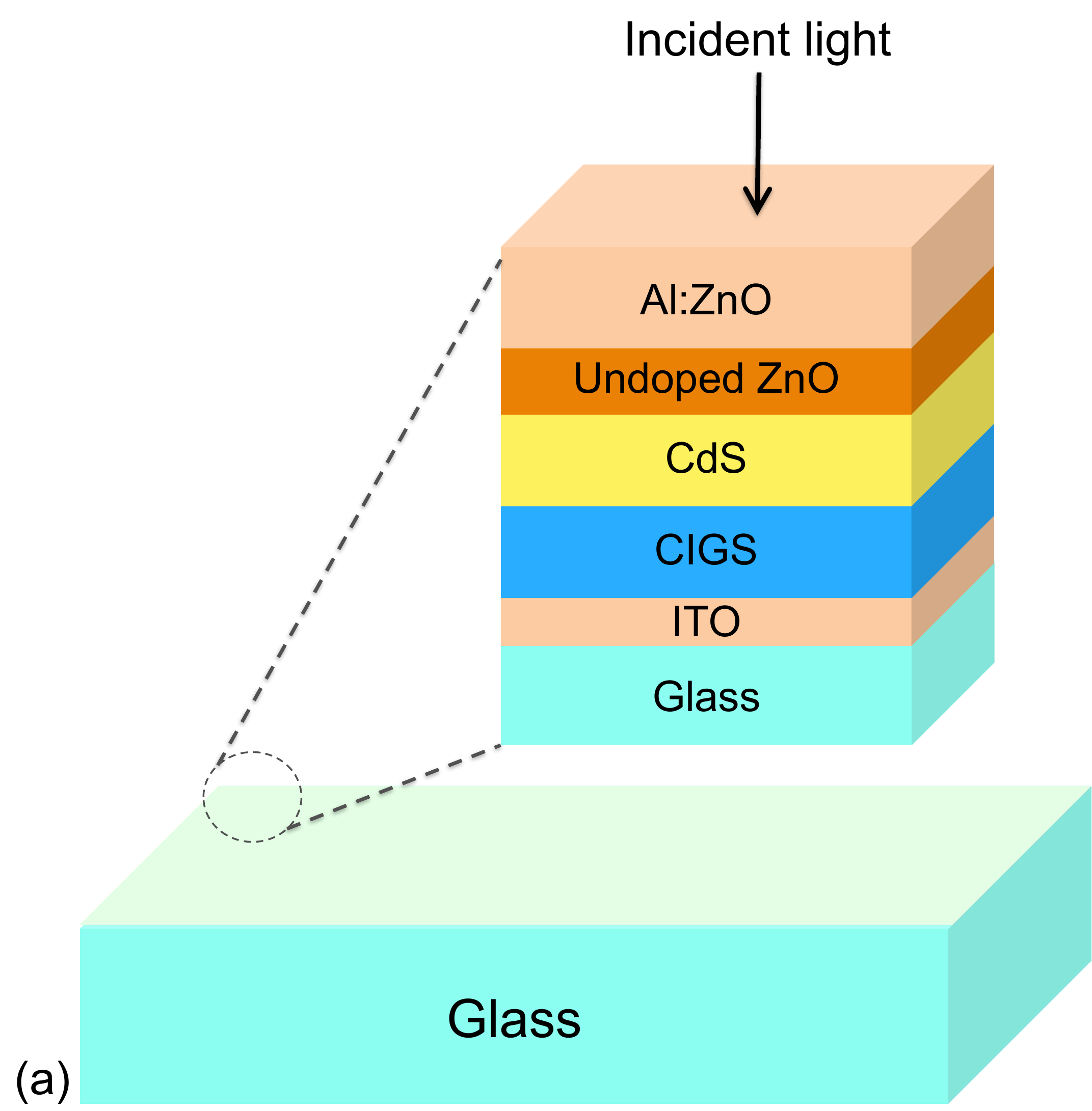}\quad
\includegraphics[width=2.7in]{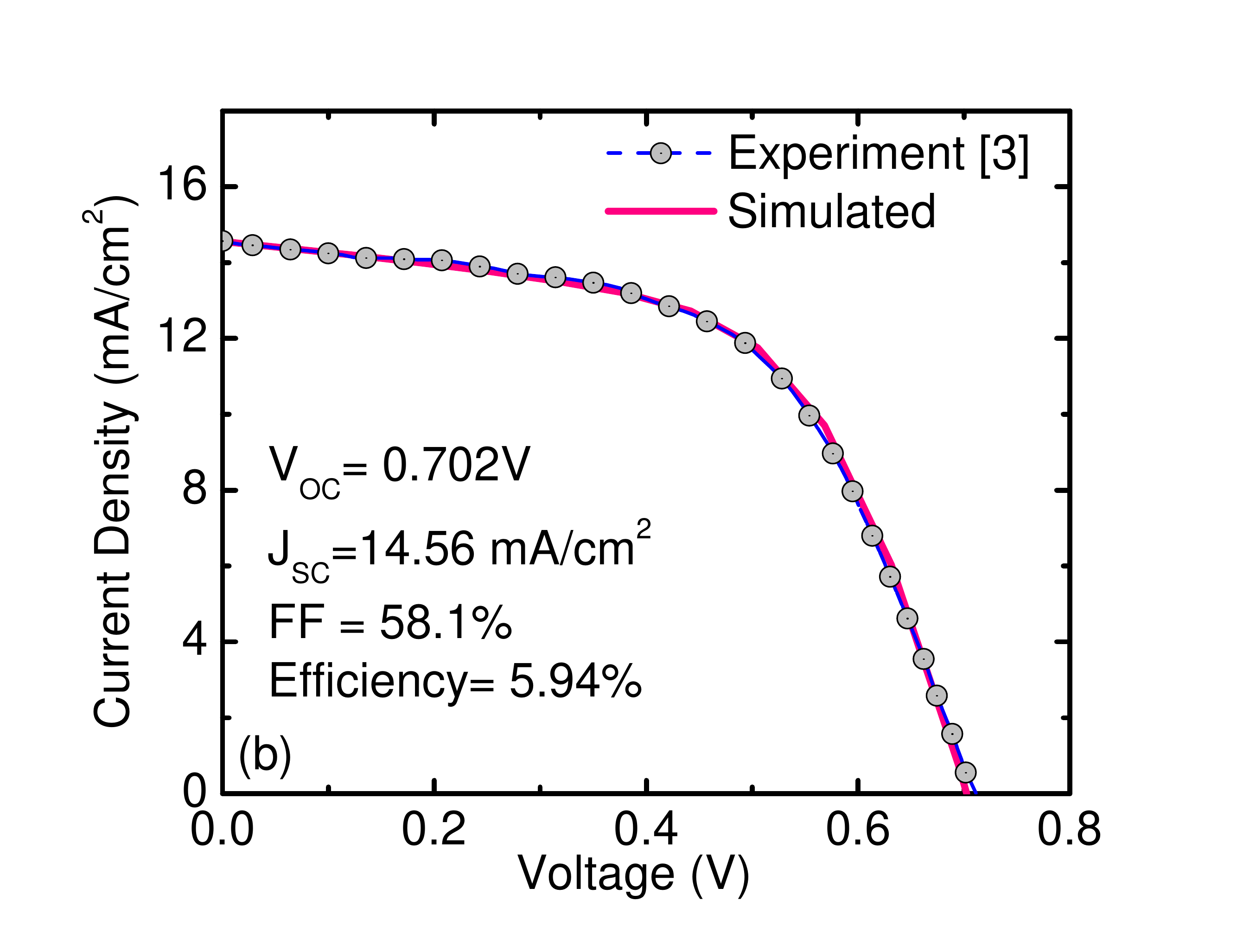}\quad
\caption{(a) Simulation of the CIGS based STPV to compare our model with experiment \cite{saifullah2016development}. Thicknesses of different layers, except the absorber layer which needs to be thin to become transparent, are similar to a conventional thin-film CIGS solar cell. (b) The model matches well with the experimental $J$-$V$ for the parameters shown in Table I. Transport is dominated by high bulk recombination and significant shunt conduction.} \label{fig2}
\end{figure}
%
\begin{figure}[ht!] 
\centering
\includegraphics[width=2.7in]{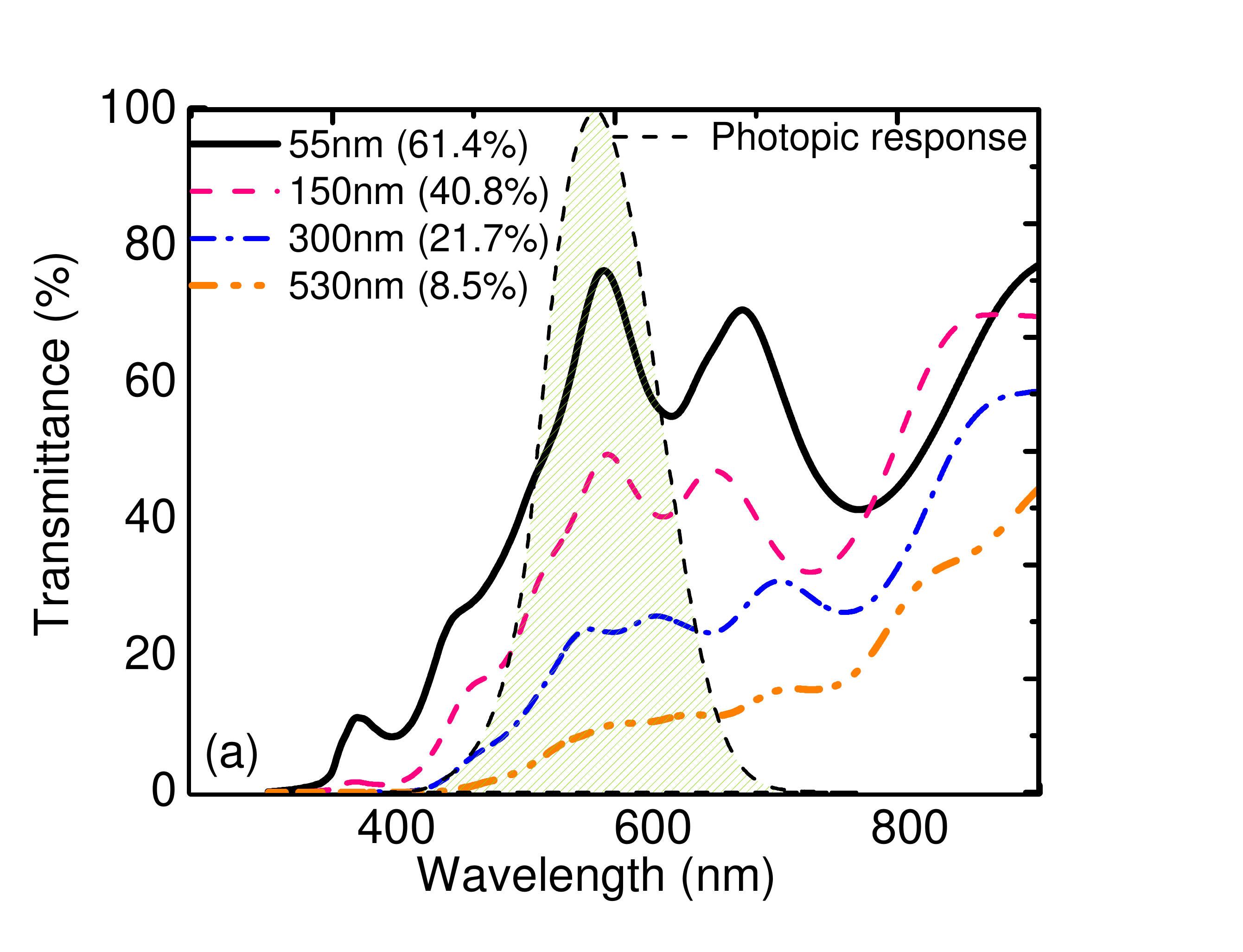}\quad
{\includegraphics[width=2.7in]{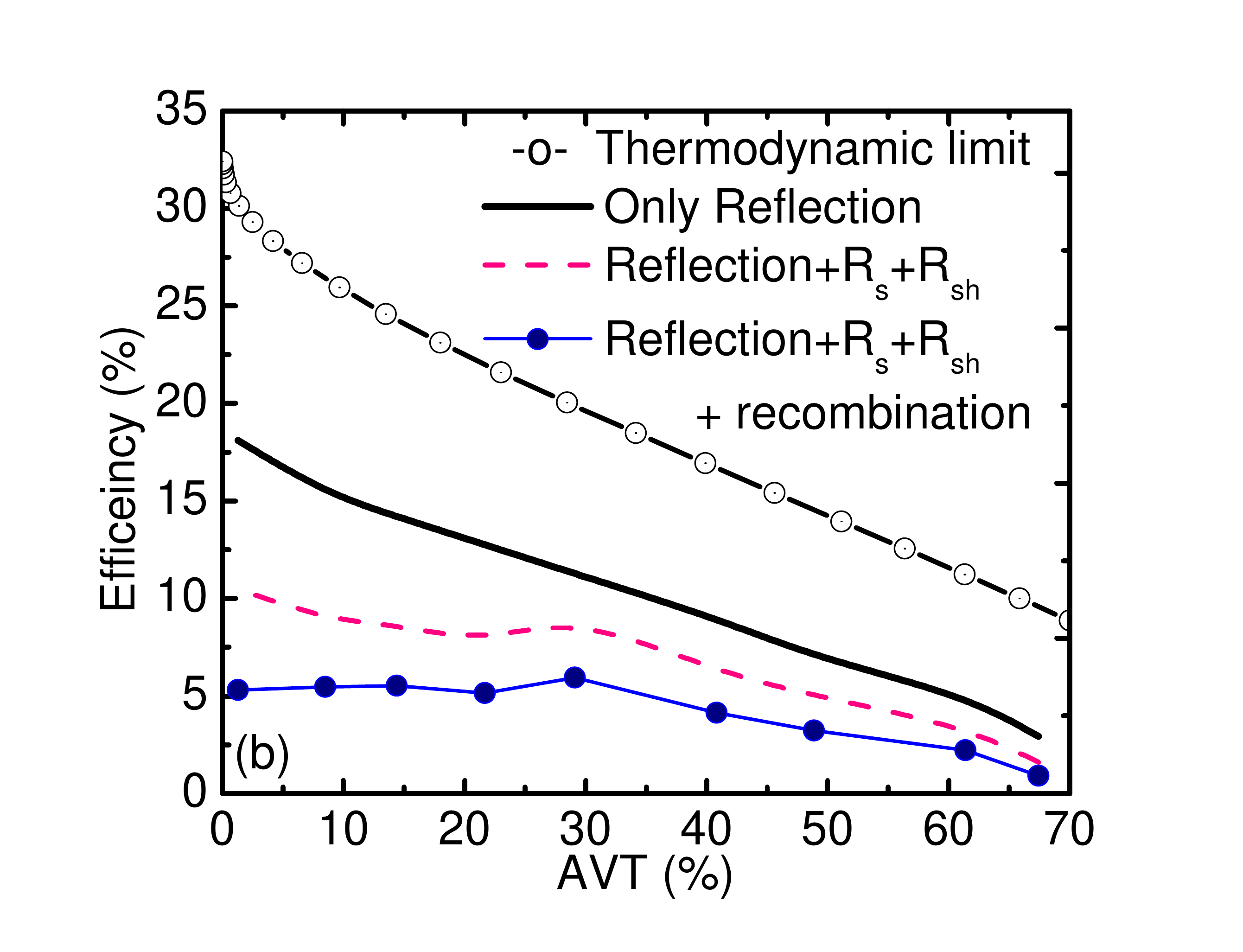}\quad}
\caption{(a) Transmission through the device for the four thicknesses (same as Fig. 1a) of absorber layer. Transmission has reduced from Fig. \ref{fig1}a due to reflection and absorption in different layers. (b) Practical efficiency at different $AVT$ as different non-idealities are introduced in the simulation. 
}
\label{fig3}
\end{figure}
In this section, we simulate the complete thin-film cell (Fig. \ref{fig2}a) comprising the front, back electrodes and buffer layers and compare our model with a recently published work on CIGS based STPV \cite{saifullah2016development}. An Indium Tin Oxide (ITO) coated glass is used as a substrate and transparent back electrode. Cadmium sulfide, undoped zinc oxide and aluminum doped zinc oxide are used as the buffer layer, window layer and top electrode respectively. Parameters of various layers are shown in Table 1. 

\begin{table*}[]
\centering
\caption{Material parameters used for simulation \cite{gloeckler2003numerical, pettersson2011baseline}}
\begin{tabular}{|l|l|l|l|l|}
\hline
\textbf{Parameters}                                                                   & \textbf{Al doped} \textbf{ZnO}                                                                              & \textbf{i-ZnO                                                                              } & \textbf{n-CdS}                                                                                     & \textbf{p-CIGS}                                                                                    \\ \hline
Thickness(nm)                                                                   & 350 & 50                                                                             & 60                                                                                    & Variable \\ \hline
Bandgap(eV)                                                                     & 3.3 & 3.3                                                                          & 2.4 & 1.5                                                                               \\ \hline
Permittivity                                                                    & 9 & 9 & 10                                                                                    & 13.6                                                                               \\ \hline
Electron effective mass                                                          & 0.2m$_0$                                                                       & 0.2m$_0$  & 0.2m$_0$                                                                    & 0.2m$_0$  \\ \hline
Hole effective mass                                                          & 0.8m$_0$                                                                        & 0.8m$_0$                                                               & 0.8m$_0$  & 0.8m$_0$  \\ \hline
Mobility, e (cm$^2$/Vs)                                                         & 100 & 100                                                                      & 100 & 100                                                                              \\ \hline
Mobility, h (cm$^2$/Vs)                                                         & 25 & 25                                                                            & 25 & 25                                                                                 \\ \hline
Electron effective density of states, $N_c$ (cm$^{-3}$)                                                         & 2.24$\times 10^{18}$ & 2.24$\times 10^{18}$& 2.24$\times 10^{18}$& 2.24$\times 10^{18}$ \\ \hline
Hole effective density of states, $N_v$ (cm$^{-3}$)                                                         & 1.8$\times 10^{19}$ & 1.8$\times 10^{19}$& 1.8$\times 10^{19}$& 1.8$\times 10^{19}$ \\ \hline
Doping concentration (cm$^{-3}$)                                                         & \begin{tabular}[c]{@{}l@{}}n: 1$\times$10$^{18}$\\ \end{tabular} & 0                                                                                & \begin{tabular}[c]{@{}l@{}}n: 2.5$\times$10$^{17}$\\ \end{tabular} & \begin{tabular}[c]{@{}l@{}}p: 1.5$\times$10$^{16}$ \footnote{Used as a fitting parameter}\\ \end{tabular} \\ \hline
SRH carrier lifetime, $\tau_\mathrm{SRH}$ (ns)                                                       & \begin{tabular}[c]{@{}l@{}}e: 10$^{-3}$, h: 0.1\\ \end{tabular}             & \begin{tabular}[c]{@{}l@{}}e: 10$^{-3}$, h: 0.1\\ \end{tabular}       & \begin{tabular}[c]{@{}l@{}}e: 0.1, h: 0.1\\ \end{tabular}                   & \begin{tabular}[c]{@{}l@{}}e: 0.1, h: 0.1\\ \end{tabular}      \\ \hline
\begin{tabular}[c]{@{}l@{}}Radiative ehp capture\\ rate (cm$^3$/s)\end{tabular} & \begin{tabular}[c]{@{}l@{}}1$\times$10$^{-10}$\\ \end{tabular}       & \begin{tabular}[c]{@{}l@{}}1$\times$10$^{-10}$\\ \end{tabular} & \begin{tabular}[c]{@{}l@{}}1$\times$10$^{-10}$\\ \end{tabular}       & \begin{tabular}[c]{@{}l@{}}1$\times$10$^{-10}$\\ \end{tabular}       \\ \hline
Metal grid: Al                                                            & \multicolumn{4}{l|}{Work function: 4.26 eV}                                                                                                                                                                                                                                                                                                                     \\ \hline
Bottom contact: ITO                                                     & \multicolumn{4}{l|}{Work function: 4.5 eV}                                                                                                                                                                                                                                                                                                                       \\ \hline
\end{tabular}
\end{table*}\label{table1}

In Ref. \cite{saifullah2016development}, the authors reported CIGS solar cells of varying absorber thicknesses, in the range of 230-500 nm, fabricated using single stage thermal evaporation method.  In order to understand the carrier transport mechanism in STPV, we fit the $J$-$V$ of their best reported device, which had CIGS thickness of 230 nm, $J_\mathrm{sc}$ = 14.57 mA/cm$^2$ and $V_\mathrm{oc}$ = 0.71 V. For 25.5\% transparency, reported efficiency of the device (5.94\%) is about a quarter of the thermodynamic limit ($>$20\% Fig. \ref{fig1}b). Fig. \ref{fig2}b shows $J$-$V$ characteristics from our simulation on top of the experimental data. The fitting procedure is as following. Since $V_\mathrm{oc}$ depends on $\tau_\mathrm{SRH}$ via the diode reverse saturation current ($J_0$), it is matched by changing $\tau_\mathrm{SRH}$. The CIGS doping concentration is then changed to match the short-circuit current, $J_\mathrm{sc}$. The best match is found for $N_A = 1.5\times10^{16}$/cm$^{3}$. The shape and the slope of the $J$-$V$ characteristics are then matched with series resistance, $R_s = 9$ $\Omega$-$\mathrm{cm}^2$ and shunt resistance, $R_\mathrm{sh} = 400$ $\Omega$-$\mathrm{cm}^2$ yielding fill factor, $FF = 58.1\%$. 

One of the dominant loss mechanisms in STPV is the trap assisted recombination of minority carriers. This is manifested through a low open-circuit voltage, even though a wide band-gap CIGS ($E_g = 1.5$ eV) was deposited. 
 As we argue later, the effect of back surface recombination is week in this kind of devices. 
  Therefore, the major non-idealities that remain are the bulk and CIGS-CdS interface recombinations. As we discuss in section VI, interface recombination alone cannot explain the low $V_\mathrm{oc}$. To match the experimental value of $V_\mathrm{oc}$, minority carrier lifetime due to trap assisted recombination is assumed as $\tau_\mathrm{SRH} = 0.1$ ns, a much smaller value than typical ($\mu$m thick) CIGS devices (200 ns reported in Ref. \cite{repins2009required}). A number of factors could contribute to the high trap density in the ultra-thin-film. Most of the defects and dangling bonds are near the interfaces, increasing the bulk surface to film volume ratio and the effective bulk recombination centers. SEM images in Ref. \cite{saifullah2016development} show that the average grain size decreases significantly with reduced film thickness (grain size varies on average between 80 nm to 490 nm for CIGS thickness of 230 to 530 nm). Reduction of grain size increases the number of grain boundaries and possible recombination centers. 
Additionally, it has previously been reported that a reduction in single-stage thermal vapor deposited chalcogenide film thickness down to $\sim$ 240 nm is associated with a noticeable deviation in molar stoichiometry and a decrease in width of the Volmer-Weber type columnar grains, both of which could lead to a considerable rise in bulk defect density in the crystal lattice \cite{zubair2019thickness}. 

The dynamics of carrier transport in the device can be understood by the double diode model of solar cell as following. According to this model, the photo-current source is connected with two diodes in parallel representing bulk recombination ($J_\mathrm{r,bulk}$) and minority carrier diffusion ($J_\mathrm{diode}$) currents. Their current contributions can be quantified by the following (under open-circuit conditions)
%
\begin{eqnarray}
J_\mathrm{L} =& J_\mathrm{diode}+J_\mathrm{r, bulk}+J_\mathrm{sh}\\
J_\mathrm{diode} \approx&\overbrace{q\frac{n_i^2}{N_A}\frac{D_n}{L_n}}^{J_{01}}\Big[\mathrm{exp}\Big(\frac{qV_\mathrm{oc}}{k_BT}-1\Big)\Big]\\
J_\mathrm{r, bulk} =& \overbrace{\frac{q\pi n_i}{2\sqrt{\tau_n\tau_p}}\frac{k_BT}{\varepsilon_\mathrm{max}}}^{J_{02}}\Big[\mathrm{exp}\Big(\frac{qV_\mathrm{oc}}{2k_BT}-1\Big)\Big]\label{eq:bulk}
\end{eqnarray}where $D_n = \mu_n k_BT/q$ and $L_n = \sqrt{D_n\tau_n}$ are the electron diffusion coefficient and diffusion length in $p$-CIGS. $n_i$ is the intrinsic carrier concentration of CIGS. The depletion region for the $p$-$n^+$ junction lies almost entirely in the CIGS side with maximum electric field, $\varepsilon_\mathrm{max}$. Using $\tau_{n,p} = \Big(1/\tau_\mathrm{SRH}+1/\tau_\mathrm{rad}\Big)^{-1}$, it can be found that $J_{02} = 1.1\times10^{-5}$ mA/cm$^2$ is 9 orders of magnitude higher than $J_{01} = 5.82\times10^{-14}$ mA/cm$^2$. Even with higher ideality factor (2), the recombination current dominates the dark current at high bias for such low, trap-dominated carrier lifetime. At this limit, the double diode model reduces to single diode model with $V_\mathrm{oc}\approx 2\big({k_BT}/{q}\big)\,\,\mathrm{ln}\big({J_\mathrm{sc}}/{J_\mathrm{02}}+1\big) \approx 0.71 $ V (same as found in the numerical simulation). The low shunt resistance results from the thin absorber layer, as discussed in Section VI, creating conductive paths between front and back electrodes. The high series resistance is likely to originate from the same mechanisms that contributed to low recombination lifetime such as small grain size leading to numerous grain boundaries. This can result in short mean free path and high series resistance. Further analysis is provided in Section VI. 

Fig. \ref{fig3}a shows the transmission through the device for different CIGS thicknesses. For comparison, we simulate the same CIGS thicknesses as done in Fig. \ref{fig1}a. Transmission has reduced (compared to Fig. \ref{fig1}a) due to 1) reflection from the air-AZO interface, 2) partial absorption of light in the auxiliary layers, 3) multiple transmissions and reflections in various layers of the device. The transmission has the signatures of a non-wavelength selective STPV, where the higher energy photons are more likely to be absorbed and the lower energy photons are transmitted. As a result the coated glass will have a brown appearance \cite{bu2015semitransparent} at any given angle of incidence. Fig. \ref{fig3}b shows the variation in cell efficiency as $AVT$ is changed. Efficiency has reduced from the thermodynamic limit due to 1) reflection from air-AZO interface, 2) series and shunt resistance, 3) non-radiative recombination. In order to understand the contribution of each loss mechanism at a given $AVT$, we incorporate them gradually in the simulation. When all these non-idealities are combined, efficiency drops by 70\% at $AVT = 30\%$ and 85\% at $AVT = 50\%$ for the parameters that matched the  experiment in Ref. \cite{saifullah2016development}. All parameters used in the simulation are shown in Table I. 

\section{Carrier transport in ultra thin-film vs thin-film solar cells}
\begin{figure}[ht!]
\centering
\includegraphics[width=2.7in]{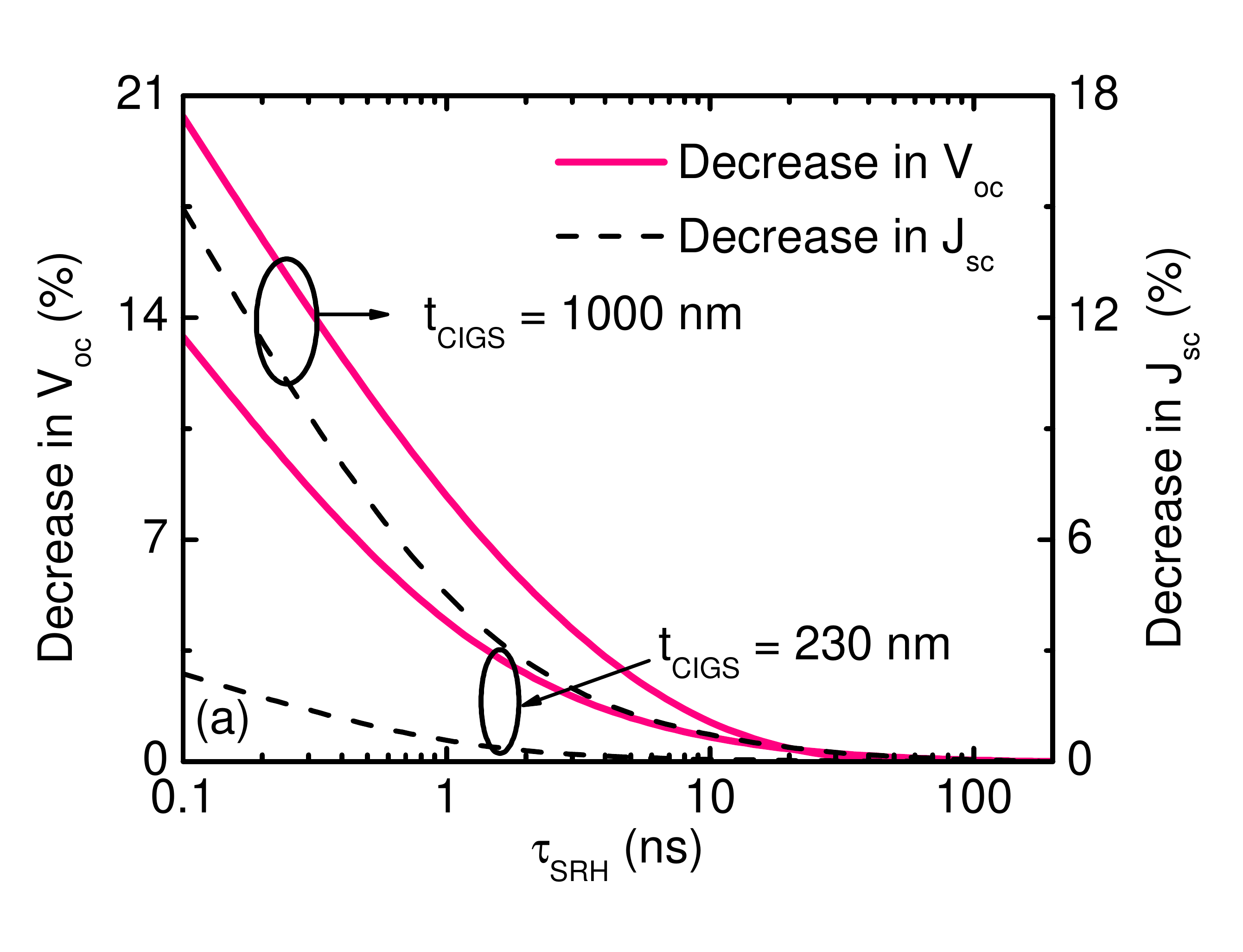}\quad
\subfigure{\includegraphics[width=2.7in]{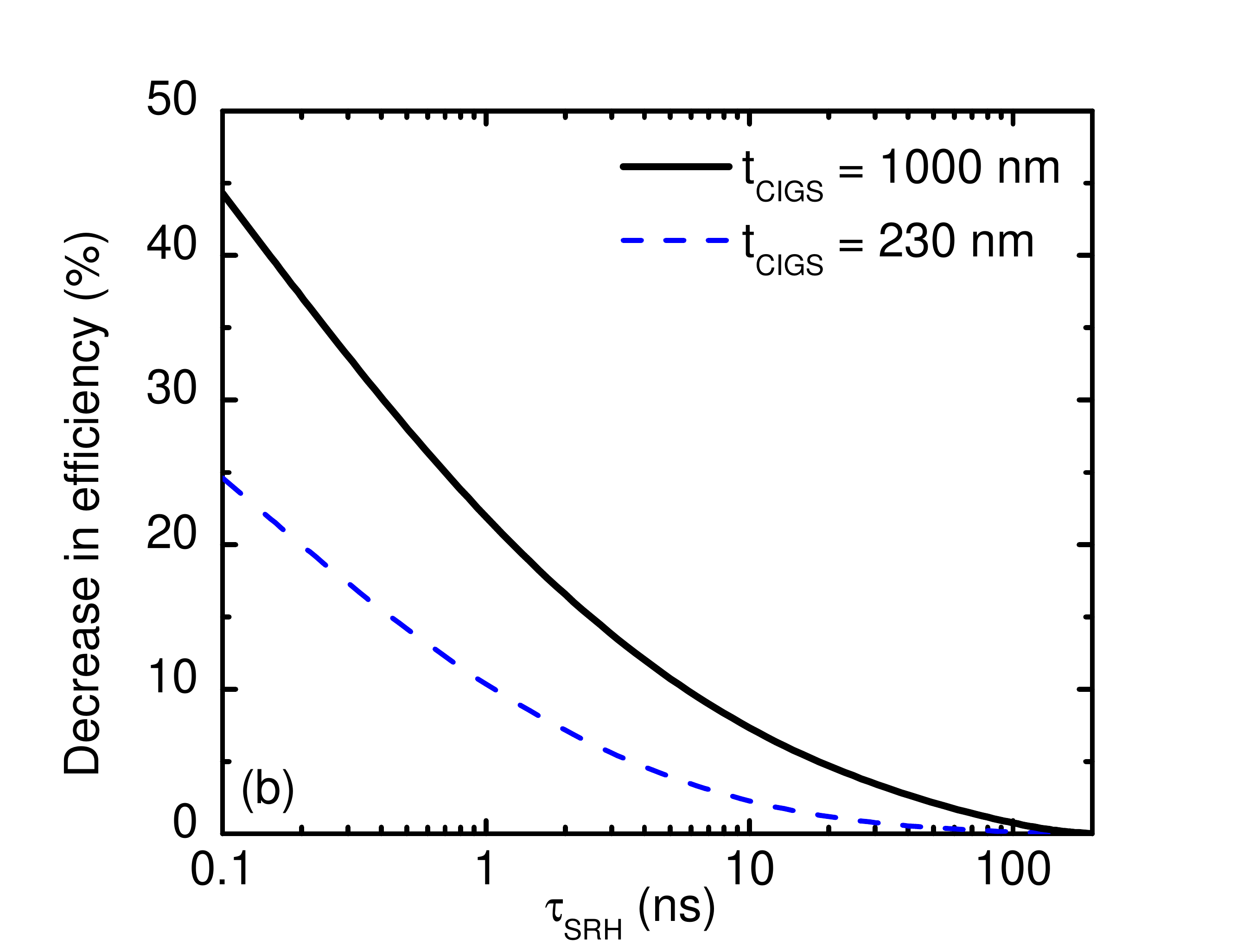}\quad}
\caption{(a) Decrease in $J_\mathrm{sc}$ and $V_\mathrm{oc}$ as bulk recombination lifetime $\tau_\mathrm{SRH}$ is changed from 200 ns to 0.1 ns for two different CIGS thicknesses. The thinner device $J_\mathrm{sc}$ and efficiency are significantly insensitive to $\tau_\mathrm{SRH}$ compared to the thicker device. $V_\mathrm{oc}$ changes more strongly with $\tau_\mathrm{SRH}$ which affect the diode reverse saturation current. (b) Efficiency of the thinner device is much less sensitive to the change in $\tau_\mathrm{SRH}$. }  \label{fig4}
\end{figure}
Thicknesses of typical CIGS based thin-film photovoltaic devices are in the range of 2 $\mu$m. In order to get at least 25-50\% visible transparency, CIGS film thickness has to be scaled down to less than 250 nm. In this thickness range, the carrier transport is significantly different from conventional thin-film devices. We show in this section the dependencies of $J_\mathrm{sc}$ and $V_\mathrm{oc}$ on carrier lifetime, carrier density and surface recombination velocities. 

In Fig. \ref{fig4}, we show the change of the output parameters - $J_\mathrm{sc}$, $V_\mathrm{oc}$ and efficiency as a function of $\tau_\mathrm{SRH}$. $J_\mathrm{sc}$ depends weakly on $\tau_\mathrm{SRH}$ in the 230 nm device compared to the thin-film device (1000 nm). $J_\mathrm{sc}$ changes by less than 3\% when lifetime is changed from 200 ns to 0.1 ns when CIGS thickness is 230 nm. This is significantly higher when the CIGS thickness is 1000 nm (11\%). While the short-circuit current depends weakly on $\tau_\mathrm{SRH}$, a high defect density in the absorber does affect the open-circuit voltage $V_\mathrm{oc}$ (reduces by 14\% in 230 nm) through the diode reverse saturation current, $J_0$. 
Efficiency falls by 44\% (Fig. \ref{fig4}b) when lifetime is decreased from 200 ns to 0.1 ns for 1000 nm while for 230 nm, this decrease is only 24\%. 

The weak dependence of $J_\mathrm{sc}$ on $\tau_\mathrm{SRH}$ for the thin absorber can be understood from the individual contributions from electron and hole diffusion currents in the quasi-neutral regions and photo-current in the depletion region  \cite{sze2006physics}
\begin{eqnarray}
J_p(\lambda) &=& -qD_p\frac{dp_n}{dx}\\
J_n(\lambda) &=& qD_n\frac{dn_p}{dx}\\
J_\mathrm{dr}(\lambda) &=& q\phi_\mathrm{in}\big[1-e^{-\alpha_\mathrm{CIGS} W_D}\big]\label{eq:jdr}
\end{eqnarray}
where $\phi_\mathrm{in}$ is the incident photon flux at the edge between $n$ side and the depletion region. $J_p$ and $J_n$ are the minority carrier diffusion currents at the edge of the depletion region in the CdS and CIGS sides of the junction respectively. $J_{dr}$ accounts for the photo-current generation in the depletion region, where high electric field accelerates these carriers towards the contacts before having the time to recombine. 

\begin{figure}
\centering
\includegraphics[width=2.7in]{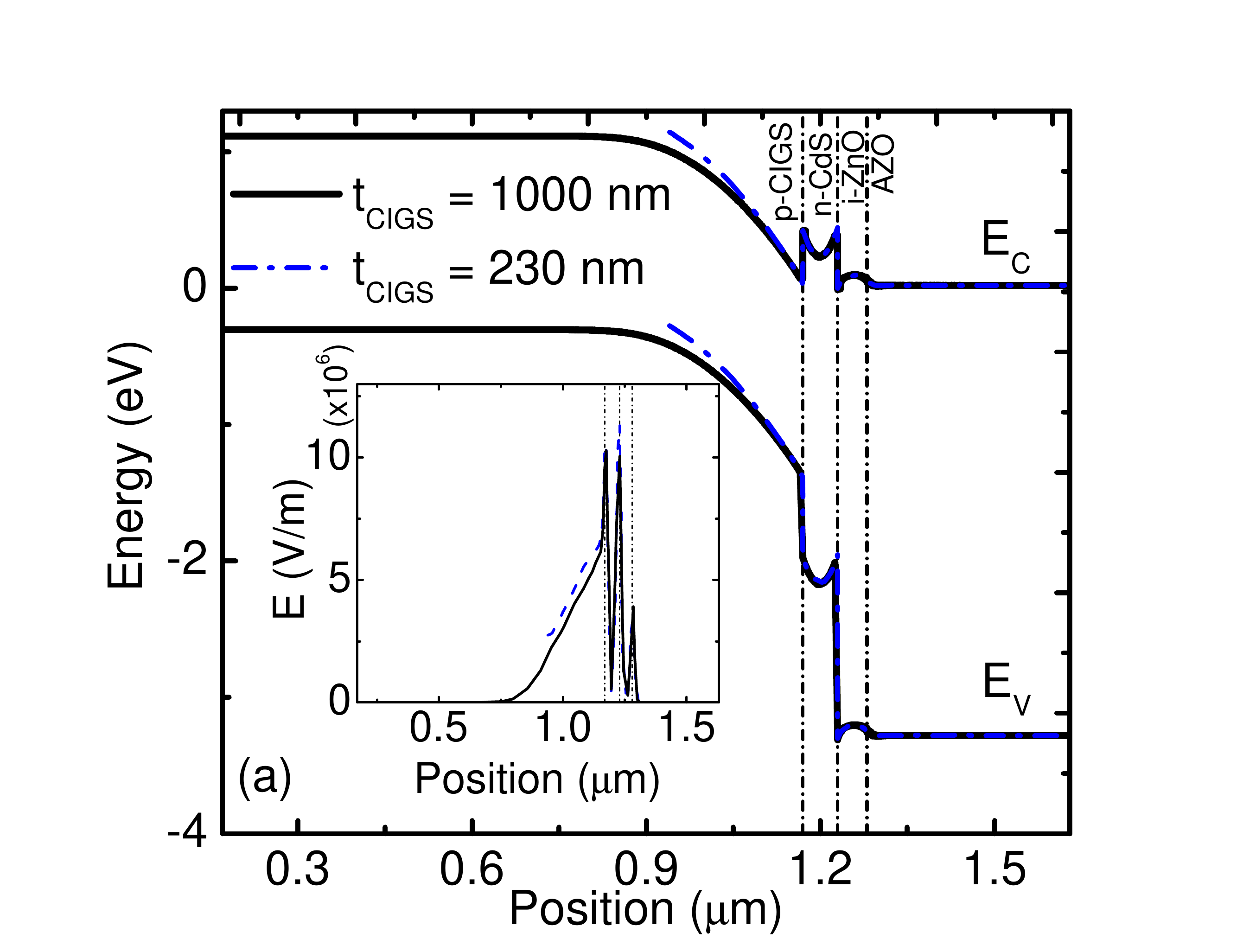}\quad
\includegraphics[width=2.7in]{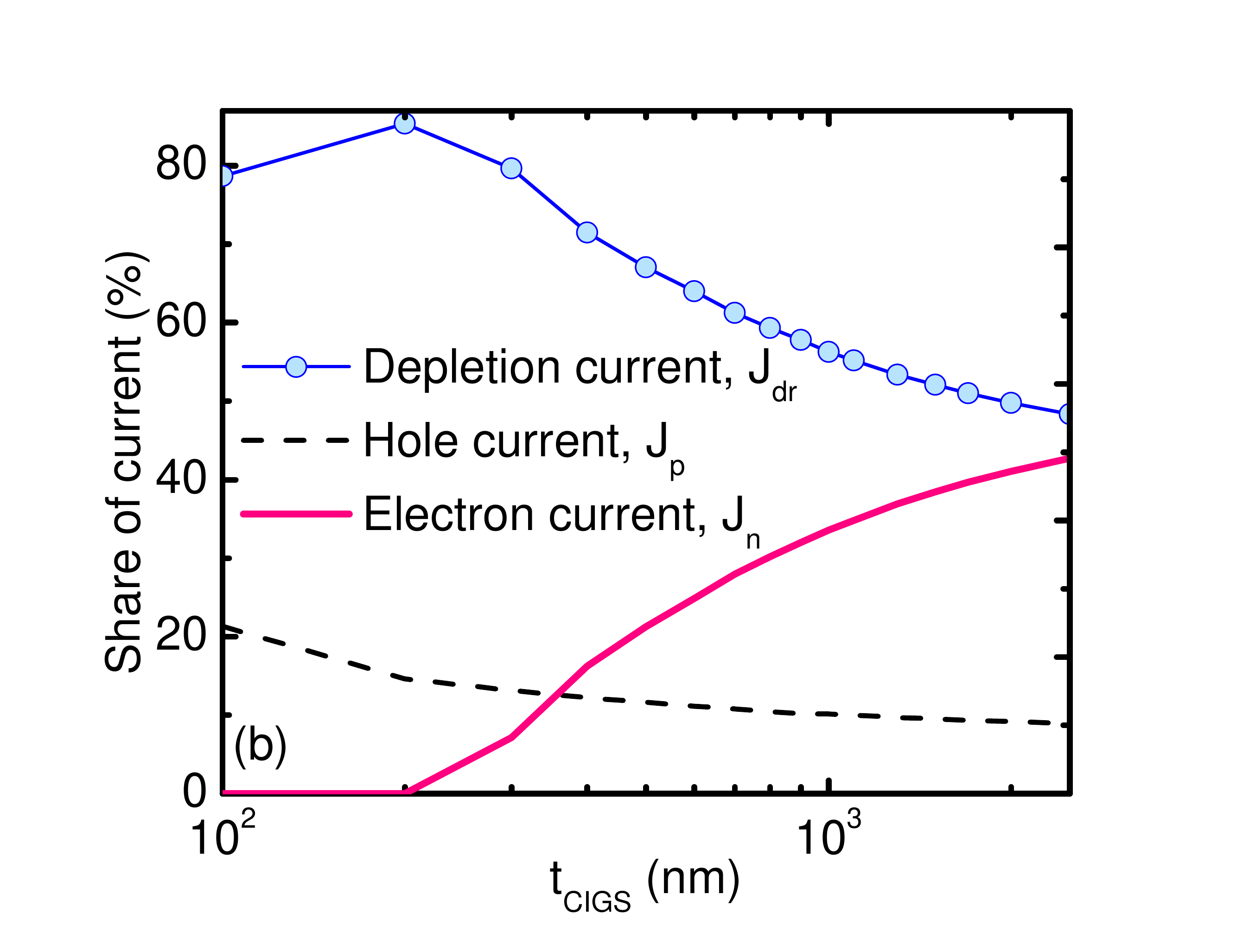}\quad
\caption{
(a) Band diagram of the device from 2D Poisson solution of the structure for two different absorber thicknesses under short-circuit condition. We find that the band alignment between CdS and CIGS is straddling or type I for both devices. The thinner device is entirely depleted and contains electric field throughout the device (inset showing electric field). (b) Share of the total current between three different current mechanisms i) hole diffusion current ii) electron diffusion current current and iii) photo-current in the depletion region. The latter carries the most current as the absorber thickness is reduced, at which limit the device is fully depleted as shown in (a). } \label{fig5}
\end{figure}

Device band diagram and electric field under short-circuit condition for two different thicknesses (230 nm and 1000 nm) are shown in Fig. \ref{fig5}a. Fig. \ref{fig5}b shows the contributions of each of these currents to the total current. For CIGS thickness less than 200 nm, more than 80\% of the current is contributed by $J_\mathrm{dr}$. For this thickness range, the $p$ side of the junction (CIGS) is entirely depleted of free carriers. This is confirmed from 2D poisson solution for device electrostatic potential and electric field (Fig. \ref{fig5}a). The carrier density used in this simulation is 1.5$\times$10$^{16}$/cm$^3$; the corresponding depletion region thickness is given by $W_D = \sqrt{2\epsilon\psi_0(1/N_A+1/N_D)} \approx 220$ nm, where $\psi_0$ is the built-in potential. Electron diffusion current starts contributing beyond this threshold, which depends on the CIGS doping density. Since the short-circuit current is dominated by photocurrent in the depletion region, where very little or no carrier recombination takes place, $J_\mathrm{sc}$ essentially becomes independent of $\tau_\mathrm{SRH}$ in this limit (also seen in Fig. \ref{fig4}a). At the thicker limit, the total current is shared between $J_\mathrm{dr}$ and $J_n$ with the latter being a function of $\tau_\mathrm{SRH}$. Because of the same reason, $J_\mathrm{sc}$ is very weakly dependent on the back surface recombination, since it does not affect $J_\mathrm{dr}$ (Eq. \ref{eq:jdr}). So we find that the short-circuit current for the ultra-thin limit depends only on two factors - 1) absorption coefficient ($\alpha$) and 2) thickness of the depletion region, $W_D$, which is a function of carrier density in the $p$-CIGS. Lower doping density in the $p$-CIGS increases $W_D$ resulting in high photo-current in the depletion region and short-circuit current. However, careful consideration is required when aiming for low carrier density since it increases series resistance.

\section{Practical limits to efficiency}
In this section, we estimate the upper limit of STPV efficiency that may be achievable in practice with better material parameters, $e.g.$ higher carrier lifetime, $\tau_\mathrm{SRH}$, lower series resistance, etc. for different $AVT$s. As we discussed, higher $V_\mathrm{oc}$ can be achieved with a higher $\tau_\mathrm{SRH}$ by increasing grain size and improving vacancy. Research on ways to improve grain morphology at this thickness range is still at the early stage;  options include using alternate transparent substrates (e.g. FTO), higher deposition temperature keeping the stoichiometry the same (and without creating additional vacancy \cite{shin2019semi}) and adopting other physical deposition methods such as sputtering. In this section, we predict the output parameters $J_\mathrm{sc}$, $V_\mathrm{oc}$, $R_\mathrm{sh}$ and $R_s$ for various scenarios of bulk and interface recombinations applied to varying CIGS thickness and gallium mole fraction. 

\subsection{Limits to short-circuit current}

For most cases, CdS thickness is much smaller than hole diffusion length ($t_\mathrm{CdS}<<L_p$). If we assume negligible front surface recombination ($S_p<10^{4}$ m/s) and $\alpha_\mathrm{CdS}L_p>>1$, we find from numerical simulation a hole diffusion current of 2.07 mA/cm$^2$ and current generated in the depletion region of 10.3 mA/cm$^2$ for CIGS thickness of 150 nm yielding $AVT$ = 40\%.  
Since CIGS is entirely depleted, there is no contribution from electron current
\begin{eqnarray}
J_n = 0
\end{eqnarray} bringing the total short-circuit current to $J_\mathrm{sc} \approx 2.07+10.3 = 12.37$ mA/cm$^2$. 
Under these assumptions, both hole current and depletion current has the form $J \propto \int q\phi\big[1-e^{-\alpha t}\big] d\lambda$, where the proportionality constant depends on the absorption and reflection through the thin-film stack, not on any carrier transport parameter. The removal of minority carrier diffusion length $L$ from the current expression completely eliminates the dependence of $J_\mathrm{sc}$ on bulk recombination via traps. 

The estimate of $J_\mathrm{sc}$ provided is the upper limit, so it is unlikely to achieve higher currents than this estimate in practice unless an anti-reflection coating is used. Possible reasons for getting lower $J_\mathrm{sc}$ in experiments than predicted here include 1) higher reflection than estimated here, 2) high front surface recombination ($S_p>10^4$ m/s) which can diminish the hole contribution and 3) lower absorption in CIGS. 

\subsection{Limits to open-circuit voltage}

Higher gallium content in CuIn$_{1-x}$Ga$_x$Se$_2$ has a higher bandgap and is expected to increase $V_\mathrm{oc}$. Previous studies reported difficulty in increasing $V_\mathrm{oc}$ beyond 0.85 V for higher bandgap CIGS absorbers \cite{huang2013origin,gloeckler2005efficiency}. In Ref. \cite{contreras2012wide}, authors fabricated solar cells with various CIGS bandgaps, up to 1.67 eV, and the highest $V_\mathrm{oc}$ was 0.829 V.  For devices with higher-quality absorber ($\tau_\mathrm{SRH}$ $>$10 ns) with back surface field to eliminate back surface recombination, the only limiting factor to $V_\mathrm{oc}$ is recombination at the CdS-CIGS interface, attributable to charged deep trap states \cite{rau2000electronic}. In this sub-section, we provide an estimate to the highest achievable $V_\mathrm{oc}$ for STPV, where the low $J_\mathrm{sc}$ is expected to degrade $V_\mathrm{oc}$ further. 

Nadenau $et.al.$  provided a trap assisted tunneling model for enhanced recombination at CIGS-CdS interface \cite{nadenau2000electronic}. According to this model, holes from CIGS tunnels into the interface, due to the high electric field at the heterostructure, and recombines producing current of the following form
\begin{eqnarray}
J_\mathrm{r,interface} = \overbrace{J_{00}\mathrm{exp}\Big(\frac{-\phi_b^p}{\xi k_BT}\Big)}^{J_\mathrm{03}}\mathrm{exp}\Big(\frac{qV}{\xi k_BT}\Big)
\end{eqnarray} Such trap assisted tunneling has been reported in the past to produce parasitic leakage paths in various semiconductor devices, $e.g.$ $p$-$n$ diodes \cite{hurkx1992new} and tunnel field effect transistors \cite{sajjad2018tunnel}. The ideality factor is given by $\xi_\mathrm{ps} = \frac{E_{00}}{k_BT}\mathrm{coth}\Big(\frac{E_{00}}{k_BT}\Big)$. $E_{00}$ is the characteristics tunneling energy given by, $E_{00} = (q\hbar/2)\sqrt{N_A/(m_h^*\epsilon)}$. $m_h^* = 0.8m$ is the hole effective mass in CIGS. The prefactor $J_{00}$ is given by
\begin{eqnarray}
J_{00} = qS_\mathrm{p, interface}N_v\frac{\sqrt{\pi q(\phi_b^p-\zeta)E_{00}}}{k_BT\mathrm{cosh}(E_{00}/k_BT)}\times\nonumber\\\mathrm{exp}\Big[\frac{\zeta}{k_BT}\Big(\frac{1}{\xi_\mathrm{ps}}-1\Big)\Big]
\end{eqnarray} where $S_\mathrm{p, interface}$ is the recombination velocity for holes at the interface, $\phi_b^p \approx 1.6 $ eV is the hole barrier (from CIGS towards the interface) under no bias condition, $\zeta = 0.13$ eV (for $N_A = 4\times10^{16}$/cm$^3$) is the difference between Fermi level and the valence band maximum on the CIGS side and $N_v$ is the effective density of states in CIGS. 
 Assuming $S_\mathrm{p,interface} = \sigma v_\mathrm{th} D_\mathrm{it} \approx 10^{5}$ m/s (for capture cross-section, $\sigma = 10^{-13}$ cm$^2$ and defect density, $D_\mathrm{it} = 10^{13}$/cm$^2$, thermal velocity, $v_\mathrm{th} = 10^5$ m/s), we find $J_\mathrm{03} = 1.5\times10^{-10}$ mA/cm$^2$ and $J_\mathrm{r,interface} \approx  9.6$ mA/cm$^2$ at $V = 0.8$ V. For an improved bulk recombination with $\tau_\mathrm{SRH}$ = 200 ns, the bulk recombination current becomes, $J_\mathrm{r,bulk} = 0.05$ mA/cm$^2$ at the same voltage (from Eq. \ref{eq:bulk}). Clearly, the interface recombination becomes the limiting factor for $V_\mathrm{oc}$ for such lifetime of bulk recombination. For $J_\mathrm{sc} = 12.4$ mA/cm$^2$ mentioned in the previous sub-section for $t_\mathrm{CIGS}$ = 150 nm, open-circuit voltage becomes, $V_\mathrm{oc} \approx \xi k_BT/q\log\Big[{J_\mathrm{sc}}/{J_\mathrm{03}}+1\Big] \approx 0.8$ V. 

\subsection{Limits to shunt and series resistance}

\subsubsection{Shunt resistance} 
Shunt paths of current conduction is the subject of numerous studies for solar cells with various models and explanations in the literature regarding their origin \cite{langenkamp2002classification}. It is a common practice to include an ohmic shunt resistance, parallel to the $p$-$n$ junction diode of the solar cell and find the output characteristics \cite{gray2011physics}. Decreasing these shunt paths (increasing resistance) is crucial to improve fill factor and energy conversion efficiency. For CIGS thin-film solar cell, several mechanisms have been proposed in the past to explain the origin of shunt paths, including, excess copper content in CIGS, microscopic pin holes creating pathways for metal diffusion from the top electrode, conductive grain boundaries etc. 
$R_\mathrm{sh}$ has strong dependence on the absorber layer thickness, $t_\mathrm{CIGS}$ with lower thickness reduces $R_\mathrm{sh}$ quickly. We have shown that in order to get at least 25\% transparency, CIGS thickness has to be as low as 230 nm. Surface roughness on both sides of CIGS layer may become comparable to absorber thickness at this range, making $R_\mathrm{sh}$ one of the limiting factors to STPV efficiency. 

Many papers reported non-ohmic behavior at low voltage dark current showing weak temperature dependence,  power-law relation with voltage and symmetric $J$-$V$ at low voltage \cite{cho2013influence, williams2015identifying}. Dongaonkar $et.al$ in Ref. \cite{dongaonkar2010universality} explained the above phenomena with the following non-ohmic shunt current model, known as the Mott-Gurney law \cite{mott1948electronic}
\begin{eqnarray}
I_\mathrm{shunt} = \mathrm{sgn}(V)A_\mathrm{sh}\frac{9\epsilon\mu}{8t_\mathrm{CIGS}^3}V^2
\end{eqnarray}where $A_\mathrm{sh}$ is the effective area where shunt conduction takes place and ``sgn" is the sign function, $\epsilon$ and $\mu$ are the electric permittivity and hole mobility of CIGS. The current is attributed to space charge limited (SCL) current, which is the dominant shunt conduction in solar cells \cite{liao2013look}. The relationship holds for high-quality material with negligible bulk defects, the limit where we would like to find the shunt resistance. The SCL current model is applicable to metal-semiconductor-metal structure where the space charge inside the semiconductor, aided by the appropriate combination of workfunctions, is set by one type of injected carriers (in this case holes). In CIGS solar cells, poor CdS and ZnO coverage resulting in Al incursion into CIGS can mimic this structure \cite{liao2013look}. Differentiating the shunt current density gives us the shunt resistance (in units of $\Omega$-device area)
\begin{eqnarray}\label{eq:rsh}
R_\mathrm{sh} = A_f\big(\frac{4}{9\epsilon\mu V}\big)t_\mathrm{CIGS}^3
\end{eqnarray}
\begin{figure}[ht!]
\centering
\includegraphics[width=2.8in]{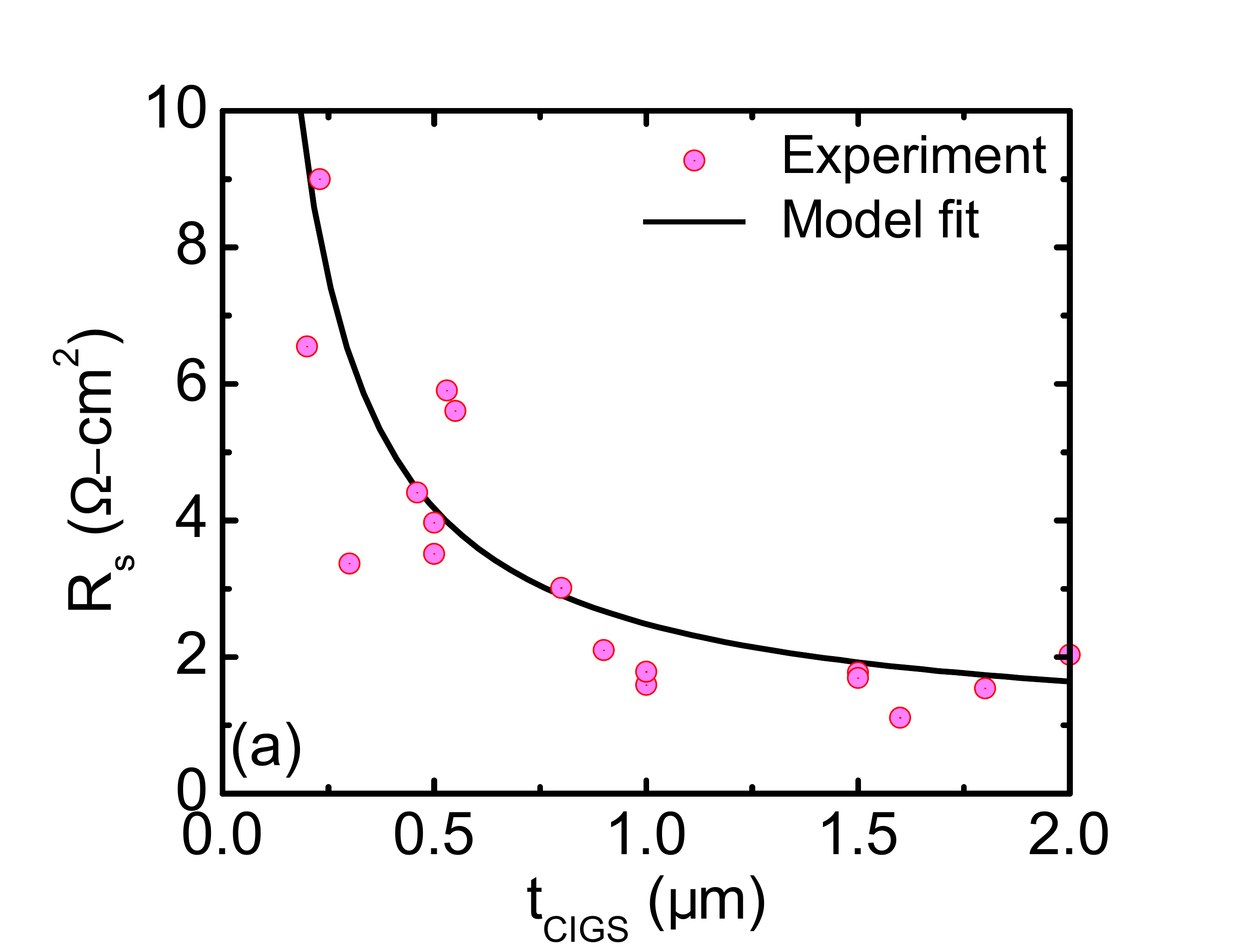}
\includegraphics[width=2.8in]{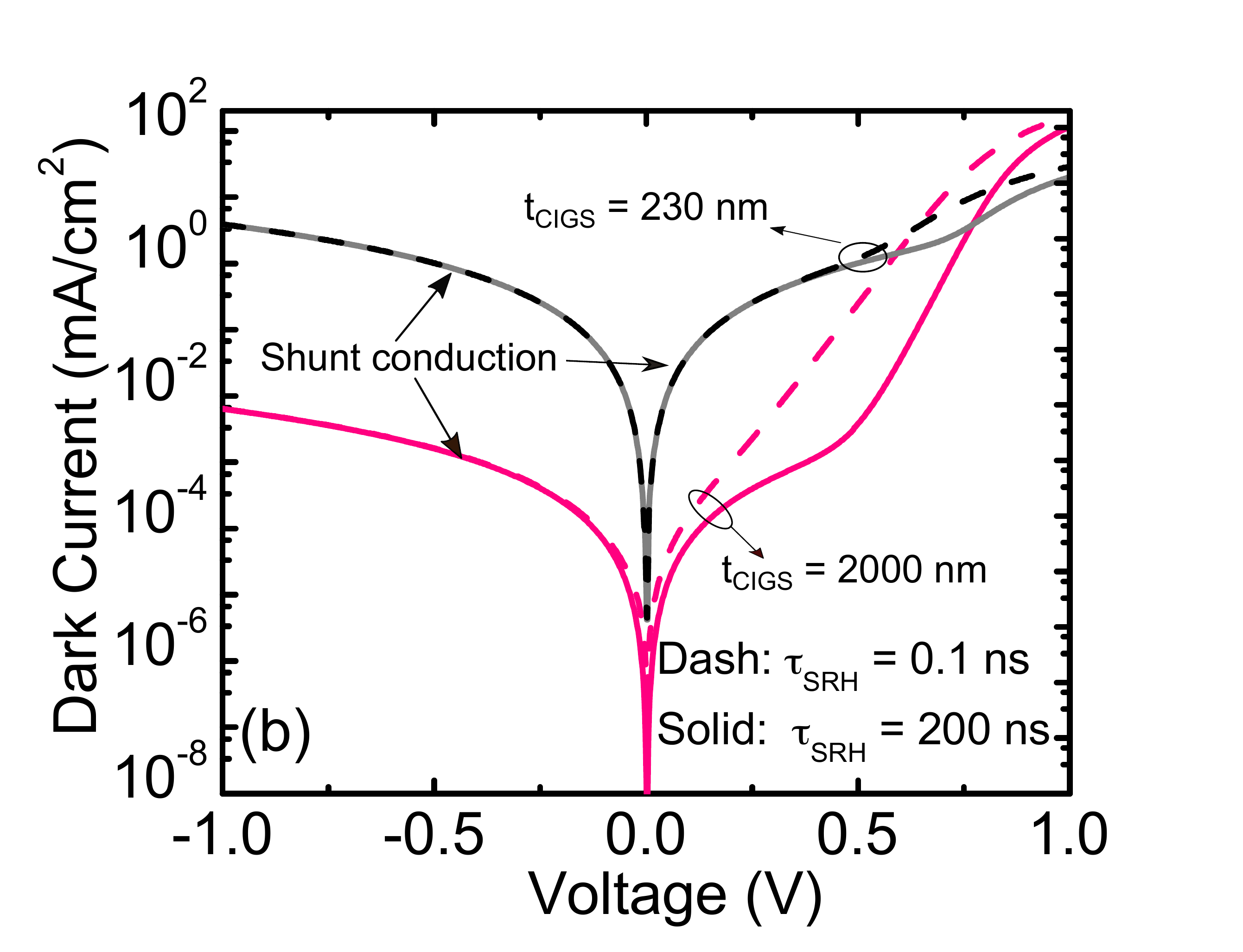}
\caption{(a) Experimental data of series resistance of CIGS thin-films as reported in the past \cite{shin2019semi, saifullah2016development, saifullah2019role, mansfield2018efficiency, kim2020flexible, park2015flower, kong2020formation, bhattacharya2013cigs, duchatelet201312, bi2017controllable, cheng2016chalcogenide, nakada2004novel} (reference order: thickness low to high) and model fit with Eq. \ref{eq:rgb}. Series resistance from experimental data is extracted by taking the slope near $V_\mathrm{oc}$. (b) Dark current for two different thicknesses, i) ultra-thin-film limit where the shunt current dominates at low bias and recombination current (either bulk or interface depending on their relative magnitudes) dominates high bias transport, ii) thicker film where recombination determines both low and high bias transport. In both cases, reverse current is determined by shunt conduction.} \label{fig6}
\end{figure}
where $A_f = A_\mathrm{CIGS}/A_\mathrm{sh}$. Device to device variability of $R_\mathrm{sh}$ makes it difficult to model and predict $R_\mathrm{sh}$. Eq. \ref{eq:rsh} gives $R_\mathrm{sh} \approx 400$ $\Omega$-cm$^2$ at V = 0.15 V for $A_f = 5\times10^5$. For a $A_\mathrm{CIGS} = 1$ cm$^2$ CIGS cell, this gives $A_\mathrm{sh} = 10^{-6}$ cm$^2$, typically found in CIGS solar cells \cite{dongaonkar2010universality}. The small effective shunt area can be justified with very localized, micrometer size non-uniformities that create small islands of shunt regions in the device. Using the same parameters, shunt resistance for a 150 nm film becomes, $R_\mathrm{sh} \approx 125$ $\Omega$-cm$^2$. at V = 0.15 V.     

\subsubsection{Series resistance} CIGS solar cells with 1.5-2 $\mu$m thickness typically report series resistance in the range of 1-2 $\Omega$-cm$^2$ \cite{cheng2016chalcogenide}. The series resistance that matched the data in Ref. \cite{saifullah2016development} for the 230 nm device was 9 $\Omega$-cm$^2$. Since CIGS makes ohmic contact with ITO \cite{nakada2004novel}, the presence of the high series resistance indicates a resistance from within the cell. When series resistances from different experiments on CIGS are plotted, a trend of increasing resistance with thinner CIGS absorber emerges (Fig. \ref{fig6}a). With shorter absorbers, the grain size decreases and the number of grain boundaries increases. This can introduce considerable resistance from grain boundary, where the trapped charges create a potential barrier \cite{seto1975electrical,lu1980quantitative}. With reduced thickness, the columnar grain boundaries, typically seen in thicker absorbers, diminish in to horizontal grain boundaries \cite{saifullah2016development, shin2019semi}. These grain boundaries, which are often parallel to the junction, are reported to enhance the series resistance manifold, especially if they are within the depletion region \cite{gloeckler2005grain}.      

Assuming grain size $L$ and potential barrier at the grain boundary $V_b$, the thermal emission current across the barrier for an applied voltage $V_a$ is given by \cite{seto1975electrical,lu1980quantitative}
\begin{eqnarray}
J_\mathrm{th} = qp\frac{1}{\sqrt{2m^*\pi}}\mathrm{exp}\big(-\frac{qV_b}{k_BT}\big)\mathrm{exp}\big[\big(-\frac{qV_a}{k_BT}\big)-1\big]\nonumber\\
\end{eqnarray} where $p$ is the carrier concentration in CIGS with effective mass $m^*$. The corresponding resistivity becomes (for small applied voltage), $\rho = \frac{\sqrt{2m^*\pi}}{Lq^2p}\mathrm{exp}\big(\frac{qV_b}{k_BT}\big)$ and the grain boundary resistance, $R_\mathrm{gb}$ becomes proportional to the number of grain boundaries in the film, $t_\mathrm{CIGS}/L$. The grain size, $L$ increases with film thickness; however, a linear dependence would result a thickness independent $R_\mathrm{gb}$. In order to match the trend seen in experiments, we use an empirical relationship $L = at_\mathrm{CIGS}^2$ and the grain boundary resistance and series resistance become as following 
\begin{eqnarray}
R_\mathrm{gb} &=& \Big[\frac{\sqrt{2m^*\pi}}{aq^2p}\mathrm{exp}\big(\frac{qV_b}{k_BT}\big)\Big]\frac{1}{t_\mathrm{CIGS}}\\
R_s &=& R_\mathrm{gb}+R_c\label{eq:rgb}
\end{eqnarray} With $a = 5\times10^3$/cm and $V_b = 0.3$ eV and adding contact resistance, $R_c =  0.8$ $\Omega$-cm$^2$ to the grain boundary resistance, Eq. \ref{eq:rgb} matches well with experimentally observed data. Contrary to the bulk resistance, which increases with film thickness, the grain boundary resistance is inversely proportional to film thickness and becomes the dominating part of the series resistance for sub-micron film thicknesses. 
\begin{figure}
\centering
\includegraphics[width=3.4in]{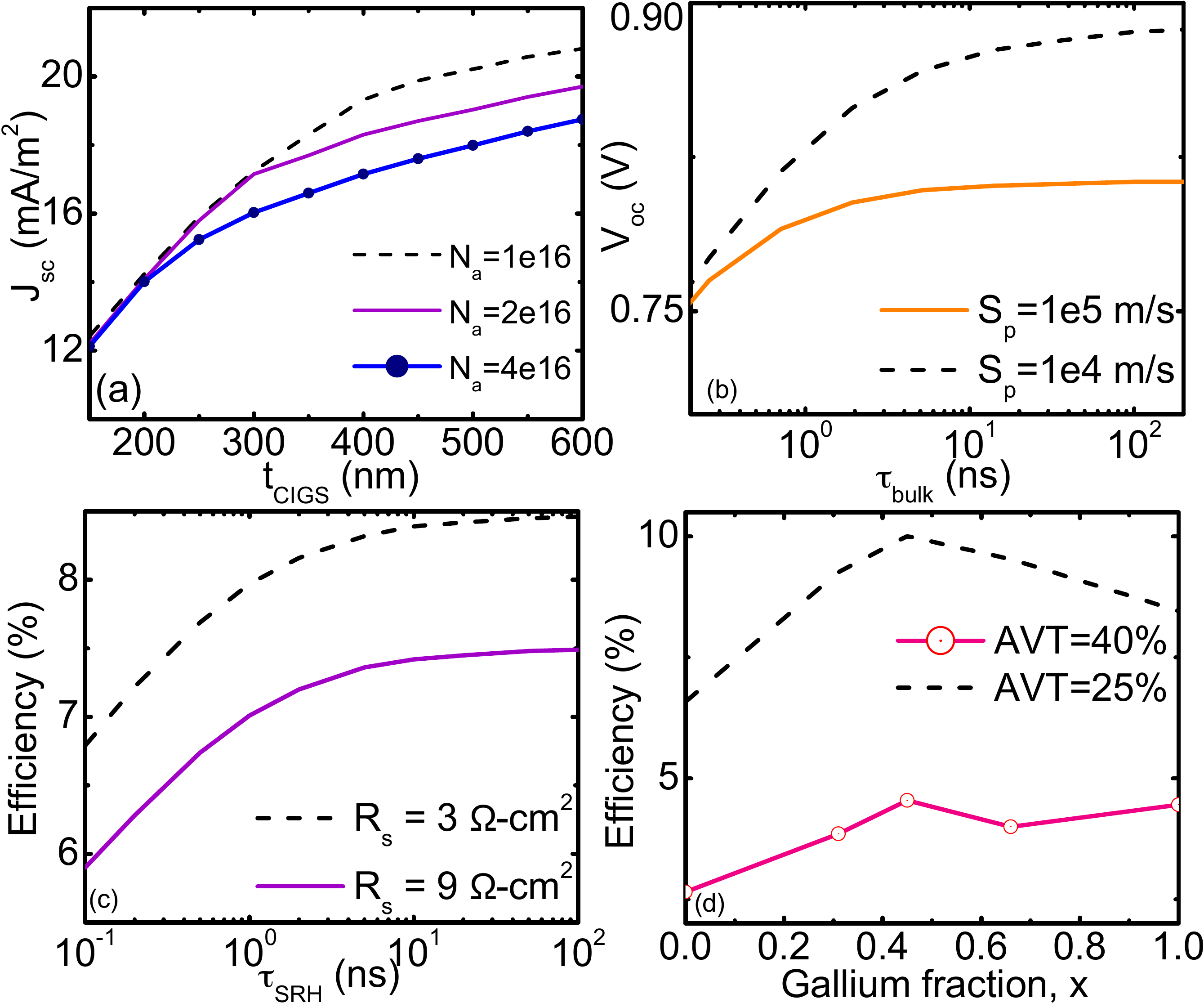}\quad
\caption{ (a) $J_\mathrm{sc}$ increases rapidly with absorber thickness $t_\mathrm{CIGS}$ and then at a slower rate when $t_\mathrm{CIGS}$ becomes larger than the depletion thickness, $W_D$. (b) Variation of $V_\mathrm{oc}$ as bulk recombination lifetime is increased. For high lifetime $\tau_\mathrm{SRH}$, $V_\mathrm{oc}$ saturates at a value dependent on the CIGS-CdS interface recombination velocity ($S_p$), which keeps $V_\mathrm{oc}$ under 0.9 V under most practical cases of CIGS solar cell. (c) Evolution of efficiency with carrier recombination lifetime and series resistance. (d) Change of efficiency as gallium mole fraction $x$ is changed. Optimum efficiency is found for $x$ = 0.27 (bandgap 1.27 eV). }
\label{fig7}
\end{figure}

Fig. \ref{fig6}b shows the combined effect of series and shunt resistance, carrier lifetime on the dark current for two different absorber thicknesses. We use series resistance of 9 and 3 $\Omega$-cm$^2$ for 230 and 2000 nm absorbers respectively. Transport at the high bias limit is determined by the combination of series resistance, bulk recombination lifetime (dashed lines) and CIGS-CdS interface recombination (solid lines). Higher series resistance leads to lower current at high bias for the thinner absorber.  Since current is determined by recombination current at high bias, open-circuit voltage is still determined by the logarithmic equation mentioned previously (section IV and VI B). For interface recombination dominated current, the $J$-$V$ is shifted to the right, which will yield higher $V_\mathrm{oc}$. The high shunt conduction for the thinner absorber plays role in low bias and in the entire reverse bias regime, resulting in symmetric power-law voltage-dependent current.  Under reverse bias for the 230 nm absorber, the current does not saturate since it is dominated by shunt conduction, giving current in the range of 0.1 - 1 mA/cm$^2$ for low reverse bias, in the same range as reported in Ref. \cite{saifullah2016development}. The thicker device has much lower shunt conduction, which resulted in the suppression of the shoulder in the low bias current. 

Based on the findings in this section, Fig. \ref{fig7} shows what to expect for $J_\mathrm{sc}$, $V_\mathrm{oc}$ and $\eta$ as the relevant parameters are changed. We find $J_\mathrm{sc}$ from the drift-diffusion solver as before. 
$J_\mathrm{sc}$ increases rapidly with absorber thickness $t_\mathrm{CIGS}$ (Fig. \ref{fig7}a) as long as the absorber is fully depleted $W_D$$>$$t_\mathrm{CIGS}$. Beyond this point, $J_\mathrm{sc}$ increases slowly due to bulk recombination ($\tau_\mathrm{SRH} = 10$ ns). With increased doping density in $p$-CIGS, a smaller $J_\mathrm{sc}$ results since the slow increase in $J_\mathrm{sc}$ kicks off at a smaller $t_\mathrm{CIGS}$ due to smaller depletion region. Fig. \ref{fig7}b shows the variation of $V_\mathrm{oc}$ with bulk recombination for two different CIGS-CdS interface recombination velocities, $S_\mathrm{p, interface}$. For high-quality CIGS absorbers ($\tau_\mathrm{SRH}> 1$ ns), the maximum achievable $V_\mathrm{oc}$ depends on $S_\mathrm{p, interface}$. Maximum $V_\mathrm{oc}$ we get is 0.81 V and 0.87 V for $S_\mathrm{p, interface} = 10^{5}$ and $10^{4}$ m/s respectively. Fig. \ref{fig7}c shows the evolution of efficiency, $\eta$ as a function of recombination lifetime, $\tau_\mathrm{SRH}$ for two different series resistances. The calculation is done for $t_\mathrm{CIGS} = 230$ nm with $AVT = $ 28\%. The efficiency is strongly influenced by series resistance at any lifetime. The experimentally observed efficiency of 5.9\% for the combination of $\tau_\mathrm{SRH}$ = 0.1 ns and $R_s =$ 9 $\Omega$-cm$^2$ is estimated to improve to $\sim$ 8.5\% for $\tau_\mathrm{SRH}$ = 200 ns and $R_s =$ 3 $\Omega$-cm$^2$. While there is no fundamental limitation to achieve these values, a systematic, targeted research is required to improve these two parameters for STPV application and reverse the trend seen in the experiments (Fig. \ref{fig6}a). In practice, the grain size will affect both series resistance and carrier lifetime simultaneously. Finally, we show the change of efficiency with gallium mole fraction ($x$) for two $AVT$s - 25\% (with $\tau_\mathrm{SRH}$ = 10 ns and $R_s =$ 3 $\Omega$-cm$^2$) and 40\% (with $\tau_\mathrm{SRH}$ = 1 ns and $R_s =$ 9 $\Omega$-cm$^2$) in Fig. \ref{fig7}d. Since bandgap and optical absorption change with $x$, different thicknesses are required to achieve the same $AVT$. The thickness dependent shunt conduction as described in Eq. \ref{eq:rsh} is included. The optimum bandgap for $AVT = 25\%$ is found for x = 0.45 (bandgap 1.27 eV yielding efficiency 10.01\%). Although the efficiency is maximum for $x = 0.45$, the widely used CuIn$_{0.7}$Ga$_{0.3}$Se$_2$ technology ($x$ = 0.3, bandgap 1.18 eV) is still a good candidate with slightly lower efficiency (9.26\%). For $AVT$ = 40\%, higher bandgap is preferred with efficiency (maximum 4.54\%) being relatively insensitive to bandgap above 1.27 eV.


Lunt $et.$ $al.$ \cite{traverse2017emergence} proposed Light Utilization Efficiency (LUE) as an indicator of how good a material is for STPV. LUE is defined as the product of energy conversion efficiency ($\eta$) and $AVT$, which means that both $\eta$ and $AVT$ should be increased as much as possible for window applications. We find that LUE (for the optimum efficiency in Fig. \ref{fig7}d) are in the range of 1.91-2.5 depending on $AVT$. This makes CIGS to have one of the highest LUE among other inorganic, non-wavelength selective STPV candidates \cite{traverse2017emergence}. The only other material that is reported to have  higher LUE is the perovskite\cite{chang2015high,kwon2016parallelized,lee2016neutral,chen2016ordered,
horantner2016shunt,lim2021semi}. 
As reported in Ref. \cite{refat2020prospect}, a 10.74\% efficient, 50\% transparent STPV based window can achieve about 45-90\% saving in building energy consumption depending on location. A lower $AVT$ of just 23\% can be used in places with abundant sunlight (such as in hot, desert regions) yielding nearly net-zero energy buildings. The practical limits of CIGS STPV efficiency reported here therefore holds promise for substantial building energy saving. Further improvement in the CIGS-CdS interface or replacing CdS with other wide bandgap semiconductors with better interface can decrease recombination current and increase $V_\mathrm{oc}$ and efficiency. A longer carrier lifetime $\tau_\mathrm{SRH}$ can be achieved by increasing grain size and reducing defects by optimizing the process parameters for the sub-500 nm thickness range. 

\section{Conclusion}
Based on numerical modeling and benchmarking with recently published experiments, we have presented carrier transport physics in CIGS-based semi-transparent photovoltaics and reported its unique characteristics compared to an opaque thin-film solar cell. We calculated the thermodynamic limits of efficiency showing its dependence on bandgap and average visible transmittance. 
The key findings can be summarized as follows: 

\begin{itemize}
\item Unlike opaque structures, the absorber in STPV is almost entirely depleted of free carriers under short-circuit condition. As a result, most of the current is contributed by carriers generated in the depletion region leading to almost negligible effects of bulk recombination on the short-circuit current $J_\mathrm{sc}$. The share of currents in the depleted region changes with CIGS thickness. 

\item This (depleted absorber) also leads to almost no effect of back-surface recombination (at the CIGS-ITO interface) on $J_\mathrm{sc}$.

\item One of the main loss mechanisms that is found while fitting the experiments is the small grain size as the absorber layer thickness is reduced. This results in short carrier recombination time and low open-circuit voltage, $V_\mathrm{oc}$.  While $V_\mathrm{oc}$ depends on bulk recombination, at the limit when bulk recombination is too strong (lifetime  $\sim$ 0.1-1 ns), the CdS-CIGS interface recombination may not have any effect. This is not the case for opaque structures since bulk recombination lifetime is typically in the range of 10-200 ns for CIGS. 

\item The reduction of grain size also results in a higher number of grain boundaries which are often parallel to the junction. This increases, as we quantify with a thermal emission model, the series resistance of the absorber layer significantly. This explains the unusually high series resistance reported in experiments with ultra-thin CIGS absorbers. 

\item Based on the space charge limited shunt current model, we find the shunt resistance to be in the range of a few hundreds of Ohm-cm$^2$, in agreement with reported experiments. 

\end{itemize}

All these carrier dynamics change with $AVT$ which requires the change of absorber thickness. The efficiency is projected for various scenarios of carrier recombination, series resistance, and gallium mole fraction $x$ in CuIn$_{1-x}$Ga$_{x}$Se$_2$. We find that the optimum composition is CuIn$_{0.55}$Ga$_{0.45}$Se$_2$ (with 10.01\% efficiency at 25\% $AVT$), whereas the widely used CuIn$_{0.7}$Ga$_{0.3}$Se$_2$ technology is still promising with 9.26\% efficiency. 


\section*{Acknowledgments}{The authors thank the support of the Department of Glass and Ceramic Engineering, Bangladesh University of Engineering and Technology (BUET). The authors acknowledge research grant from Bangladesh Energy and Power Research Council (EPRC), grant no: 58-2018-003-01.}



\bibliographystyle{apsrev4-1}
\label{bblstart}

\end{document}